\documentclass[preprint, superscriptaddress, notitlepage]{revtex4-2}
\usepackage{amsfonts}
\usepackage{amssymb}
\usepackage[english]{babel}
\usepackage{hyperref}   
\usepackage{color}
\usepackage{float}
\usepackage{placeins}
\usepackage{amsmath}
\usepackage{hyphenat}
\usepackage{graphicx}  
\usepackage{caption}
\usepackage{subcaption}
\usepackage{siunitx}
\usepackage[symbol]{footmisc}
\usepackage{natbib}

\newcommand{\bn}{\mbox{\boldmath $n$}}
\newcommand{\bN}{\mbox{\boldmath $N$}}

\newcommand{\bx}{\mbox{\boldmath $x$}}

\newcommand{\bB}{\mbox{\boldmath $B$}}
\newcommand{\br}{\mbox{\boldmath $r$}}

\newcommand{\bj}{\mbox{\boldmath $j$}}
\newcommand{\be}{\mbox{\boldmath $e$}}
\newcommand{\bv}{\mbox{\boldmath $v$}}

\newcommand{\bG}{\mbox{\boldmath $G$}}
\newcommand{\bD}{\mbox{\boldmath $D$}}

\begin{document}
\raggedbottom 
\title{Skyrmionic division}

\author{Charles Kind}
\affiliation{Department of Computer Science, University of Bristol, Bristol BS8 1UB, UK}
\author{David Foster}

\date{\today}

\begin{abstract}
Magnetic skyrmions and skyrmion bags are nano-scale spin textures whose stability, size and ease of manipulation make them strong contenders for next generation data and logic applications. Skyrmion bags are composite skyrmions of any integer topological degree and this means they have emergent properties not present in unitary degree skyrmions. Here we present models and theoretical descriptions demonstrating the process of skyrmionic division where a skyrmion or skyrmion bag can be divided into two or more skyrmionic structures. This exciting result could pave the way to entirely new types of skyrmion device in the emerging fields of spintronics and neuromorphic computing.
\end{abstract}

\maketitle
\section{Introduction}

As we reach the practical and theoretic limits of our current generation integrated circuit technologies, with the demise of both Moore's anecdotal law and Dennard's scaling law \cite{dennardDesignIonimplantedMOSFET1974} more than a decade ago, we find ourselves in need of effective alternatives for high performance computing systems. This issue is exacerbated by the trend towards highly parallel systems, such as deep neural networks, and the exploitation of huge, and growing, datasets requiring large amounts of data processing \cite{horowitzComputingEnergyProblem2014, koggeExascaleComputingTrends2013}.

Both individual skyrmions and skyrmion bags \cite{fosterTwodimensionalSkyrmionBags2019} have been proposed as a possible route to a new generation of low power, high density memory \cite{fertSkyrmionsTrack2013, zhangMagneticSkyrmionLogic2015, fosterTwodimensionalSkyrmionBags2019} and as potential candidates for use in neuromorphic computing systems \cite{chenSkyrmionicInterconnectDevice2020, songSkyrmionbasedArtificialSynapses2020,grollierNeuromorphicSpintronics2020}. 

\begin{figure}[h!]
\centering
\includegraphics[width = 0.90\columnwidth]{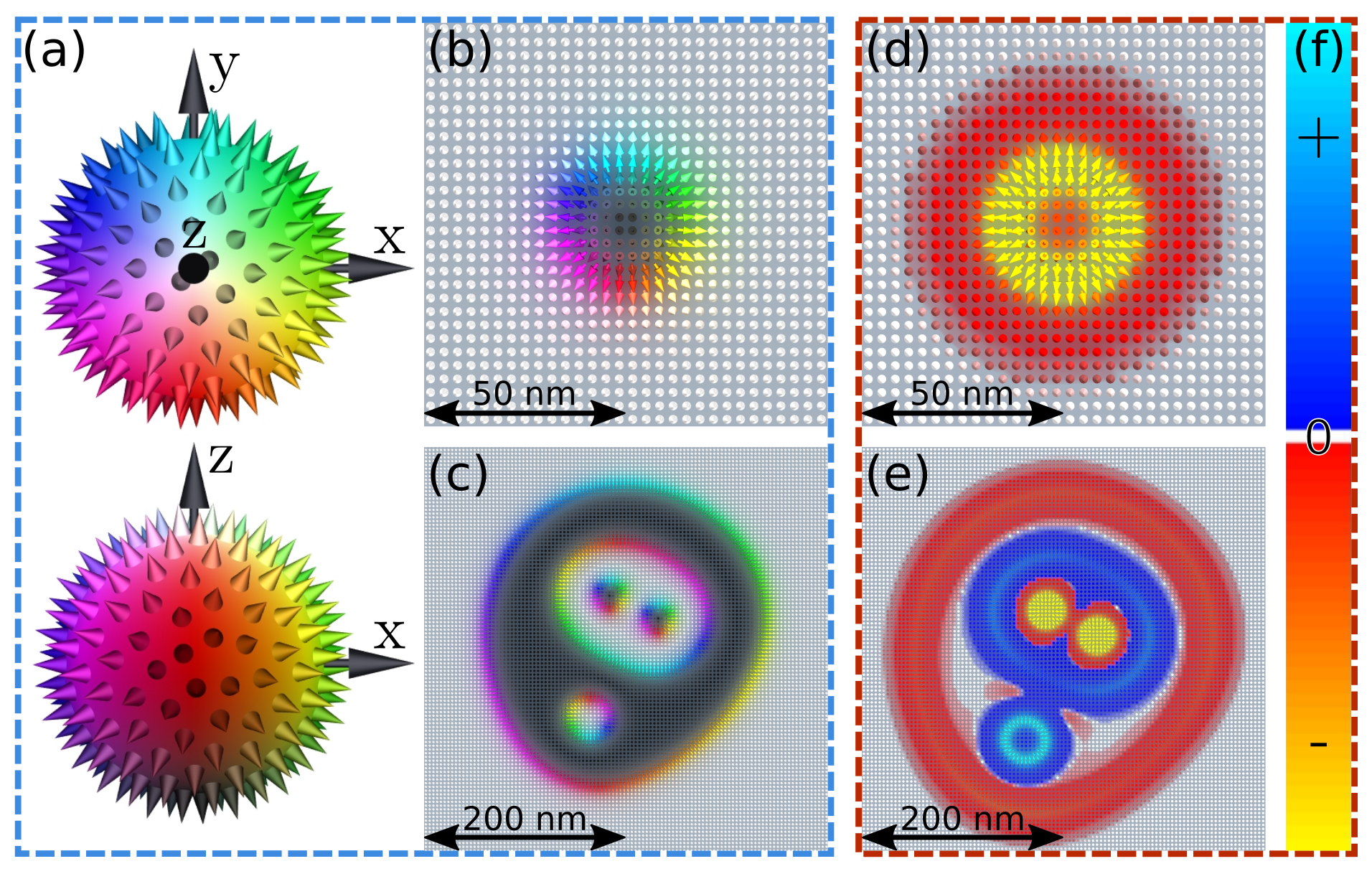}
\caption{\textbf{A skyrmion and skyrmion bag in two different colour maps.}
   \textit{	 
   \textbf{(a)} The Runge colour sphere with vectors normal to the surface representing the perfect cover of the two sphere ($S^2$) by a single skyrmion.
   \textbf{(b)} A simulated skyrmion. Note how the colours around the boundary map to the equator of the sphere in \textbf{(a)}. 
   \textbf{(c)} A simulated $S(\tilde{S}(2),\tilde{S}(0))$. 
   \textbf{(d)} The same simulated skyrmion from \textbf{(b)}. Here the colour map denotes the local topological degree density, see \textbf{(f)}.
   \textbf{(e)} The same simulated skyrmion bag from \textbf{(c)} under the topological degree density colour map.
   \textbf{(f)} The colour map for the topological degree density, $q(x)$, from Eq. (\ref{TopDeg}). Note the gap around $0$, this is to aid visualisation by making all vectors that have close to zero contribution to the topological degree coloured white. The background of the skyrmionic simulations are set to grey for contrast but this has the effect of making the white, $+e_z$, vectors appear a slight grey shade as the detail gets finer.
  }}
\label{fig1}
\end{figure}

Magnetic skyrmions are particle-like spin textures in the magnetization of chiral magnets \citep{bogdanovThermodynamicallyStableMagnetic1994}. Present in both bulk magnetic materials and interfacially in magnetic multilayers, they are stabilised by the Dzyaloshinkii–Moriya interaction (DMI) \cite{dzyaloshinskyThermodynamicTheoryWeak1958, moriyaAnisotropicSuperexchangeInteraction1960}. The magnetisation vectors of a single skyrmion, in the two dimensional (2D) continuous model, is a cover of the two-sphere, as shown in Fig. \ref{fig1}, and carries a unitary degree, $|Q|=1$, defined as
\begin{align} \label{TopDeg}
Q = \frac{1}{4 \pi} \int \bn \cdot \left( \partial_1 \bn \times \partial_2 \bn \right) \: d^2 x = \int q \left( x \right) d^2 x,
\end{align}
where $\bn \left( \bx \right)$ is the unit vector field of magnetization. Under this definition, a single stable skyrmion has the degree $Q=-1$.


Skyrmion bags \cite{fosterTwodimensionalSkyrmionBags2019}, or sacks \cite{rybakovChiralMagneticSkyrmions2019}, are nested skyrmionic structures of any integer topological degree. The textures are typically composed of a single skyrmion outer boundary and then a number of inner upside down skyrmions (UD-skyrmions) whose degree, $Q=1$, is opposite to the outer skyrmion and which can themselves contain skyrmions. We define a simple skyrmion bag as one which has a single outer skyrmion boundary and a number of inner UD-skyrmions and use the notation $S(n)$ where $n$ denotes the number of UD-skyrmions. This bag then has a total degree of $Q=n-1$. A simple skyrmion bag whose outer boundary is a UD-skyrmion, and therefore contains $Q=-1$ skyrmions, we denote as an $\tilde{S}(n)$ bag with degree $Q=1-n$. In this way we are able to label complex nested structures clearly, see Fig. \ref{fig1}. Under this notation a single skyrmion would be an $S(0)$, empty bag.

Due to their composite nature skyrmion bags possess emergent properties absent from unitary degree, $Q=|1|$, skyrmions such as varying degrees of angular deflection from applied electric current determined by the textures total topological degree \cite{kindMagneticSkyrmionBinning2021} allowing their separation.

Skyrmionic technologies are fundamentally different, in structure and implementation, from complementary metal oxide semiconductor (CMOS) ones. The particle-like nature of skyrmions allows for entirely different problem approaches such as finding the shortest path for a given graph \cite{tomaselloRoleMagneticSkyrmions2021},  logic in memory devices (LIM) \cite{gnoliSkyrmionLogicInMemoryArchitecture2021} or building artificial neurons \citep{chenNanoscaleRoomTemperatureMultilayer2020}. These differences allow for utterly new approaches to computational problems.

Here we present theory and simulation describing the division of both individual skyrmions and skyrmion bags into two or more stable skyrmions or skyrmion bags. This result opens a new door in skyrmionic technologies allowing novel spintronic solutions to utilise the magnetic field equivalent of organic cell division.

\section{Theory and initial results}

We present an overview of the collective coordinate model and governing equations, for a more detailed breakdown see our earlier paper, Kind and Foster, 2021 \cite{kindMagneticSkyrmionBinning2021}, where we demonstrate that skyrmion bags, with different topological numbers $Q$, can be separated, or binned, by the effect their topological degree has on their trajectory \cite{kindMagneticSkyrmionBinning2021}. We apply the Thiele \cite{thieleSteadyStateMotionMagnetic1973} approach to the dynamics of our skyrmionic magnetisation textures assuming the stationary limit, where the magnetisation texture moves with constant velocity, and assuming, here, that the texture does not deform. We consider the current perpendicular-to-plane (CPP) geometry using Slonczewski's \cite{slonczewskiCurrentdrivenexcitationmagnetic1996} model. The travelling wave ansatz, $\bn \left( \br , t \right) = \bn \left( \br - \bv_d t\right)$ , is then applied to the Landau-Lifshitz-Gilbert \cite{gilbertphenomenologicaltheorydamping2004} dynamical equation 
with torque to produce \cite{thiavilleMicromagneticunderstandingcurrentdriven2005a, everschorCurrentinducedrotationaltorques2011, eliasSteadymotionskyrmions2017}
Thiele's equation \cite{thieleSteadyStateMotionMagnetic1973}. 
\begin{align}\label{ThieleCPP}
\bG \times \bv - \alpha \bD \cdot \bv + \mathcal{B} \cdot \bj = 0,
\end{align}
where $\bG = 4 \pi Q \be_z$ is the gyrocoupling vector, with $Q$ from Eq. (\ref{TopDeg}), $\bD$ the dissipative tensor 
%
$\bv$ the velocity of the centre of mass of the skyrmion bag, $\bj$ the current vector and $\alpha$ the Gilbert damping constant. 
$\mathcal{B}$ is the driving force term linked to the spin transfer torque (STT). 

%





A skyrmion at rest is a critical point of the space of permissible configurations of the micromagnetic energy functional, 
\begin{align}\label{mmeng}
E(\bn)=E_{\mathrm{ex}} + E_{\mathrm{DMI}}
+ E_{\mathrm{demag}} + E_{\mathrm{anis}} + E_{\mathrm{B}},
\end{align}
%
%
%
with the energy terms being the Heisenberg exchange, antisymmetric exchange (DMI), demagnetisation, easy axis anisotropy and external magnetic field energies respectively.

Under strain skyrmions and skyrmion bags deform \cite{wangUniaxialStrainModulation2018, elhogFinslerGeometryModeling2021} and given sufficient strain can form chiral stripe domains \cite{linEdgeInstabilityChiral2016}.
Further Thiele's equation,  (\ref{ThieleCPP}), suggests that the constituent parts of a skyrmion bag that retain their low energy configuration symmetries will, independently of the entire texture, be effected locally as skyrmionic structures deflected by polarised current.

\begin{figure}[h!]
\centering
\includegraphics[width = 0.90\columnwidth]{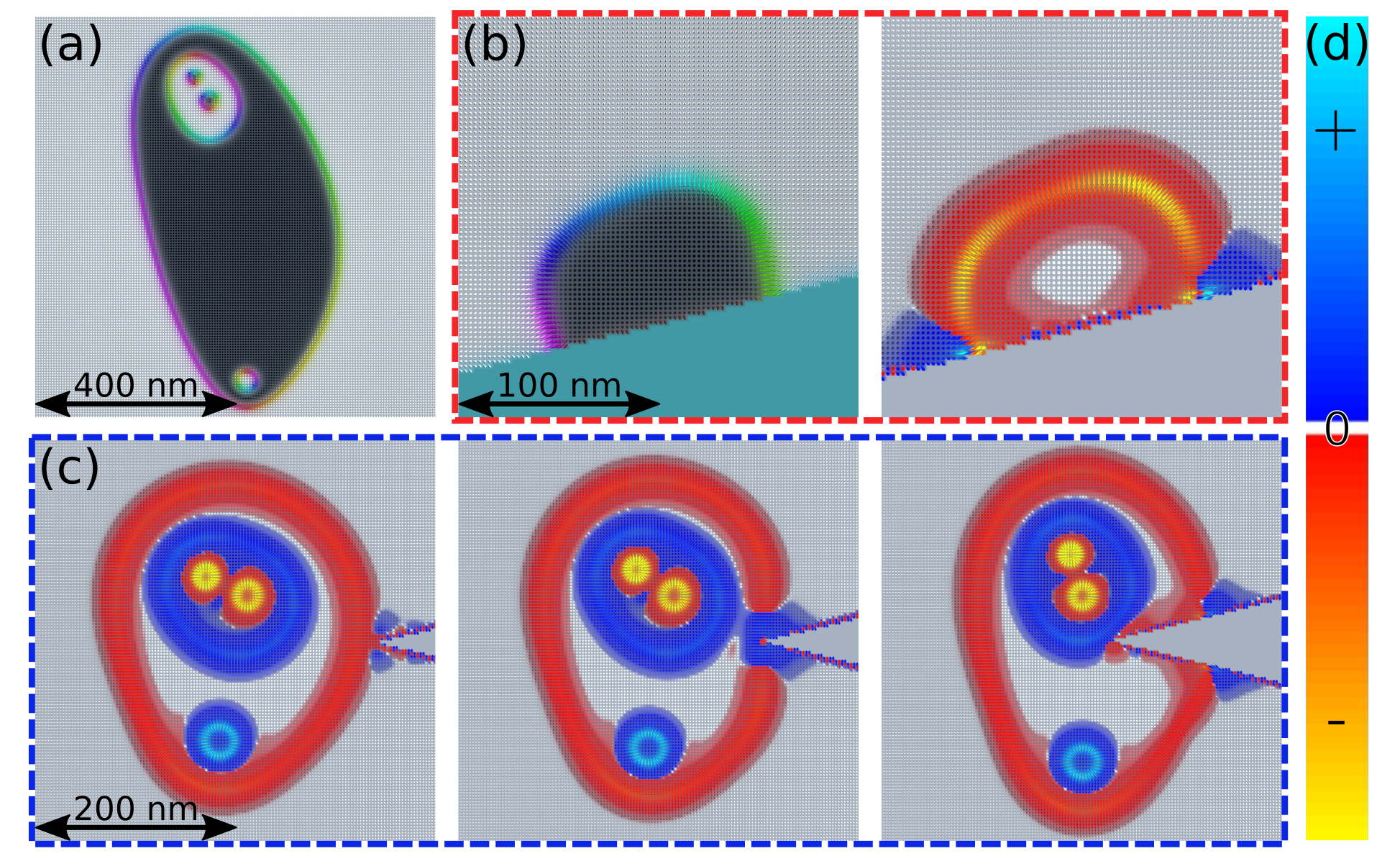}
\caption{\textbf{Stretching, crushing and cutting skyrmions and skyrmion bags.}
   \textit{
   \textbf{(a)} A simulated $S(\tilde{S}(2),\tilde{S}(0))$ bag stretching under applied polarised current.
   \textbf{(b)} A simulated skyrmion pushed onto a boundary and forming a N\'eel domain wall. Shown in both colour maps.
   \textbf{(c)} A simulated $S(\tilde{S}(2),\tilde{S}(0))$ bag being deformed by a cutting boundary. Note the formation of the opposing degree, blue positive, area at the cutting edge. Each image is $2$ ns apart.
   \textbf{(d)} Colour chart for the local topological degree $q(x)$.
  }}
\label{fig2}
\end{figure}

The upshot of this is that a skyrmion bag that contains two skyrmionic structures with opposing sign degrees $Q$, will stretch under sufficient applied current. As an example consider an $S(\tilde{S}(0),\tilde{S}(2))$ with total $Q=-1$, see Fig. \ref{fig1}(b), where the outer boundary has $Q=-1$, the inner $\tilde{S}(0)$ has $Q=1$ and the inner  $\tilde{S}(2)$ has $Q=-1$. Under simulation, this stretching is exactly what we see, Fig. \ref{fig2}(a).

The minimum energy configuration for both a skyrmion and a skyrmion bag is for the outer boundary to have axial symmetry. This axial configuration does not possess any regions of anti-topological degree, $q(x)$, see Fig.\ref{fig1}. This symmetry may be somewhat malformed in the case of more complex, nested skyrmion bags but the principle remains. When the boundary is deformed against it's natural convex curve it is energetically optimal for the skyrmion to create regions of negative topological charge density as shown in Fig. \ref{fig2}(c), where a bag is 'pushed' onto a sharp defect. As the bag continuously deforms from a circle to a figure eight type configuration, greater amounts of anti-topological degree are produced until it is energetically favourable for the texture to split and create an anti-skyrmion which shrinks to a point.

Skyrmions are initially repulsed and eventually deformed and eliminated when pushed into a boundary with sufficient force \cite{zhangSkyrmionskyrmionSkyrmionedgeRepulsions2015}. Before elimination the broken skyrmion forms a magnetic domain wall of N\'eel type terminating at either end, of the wall, on the boundary, see Fig. \ref{fig2}(b). 

We put these results and theory together and design a skyrmionic slicer in order to divide skyrmions and skyrmion bags. Below we detail the processes by which we achieved this result we then discuss why the processes took the forms they did and some theoretical implications of these results.


\section{Skyrmion Division}

\begin{figure}[h!]
\centering
\includegraphics[width = 0.90\columnwidth]{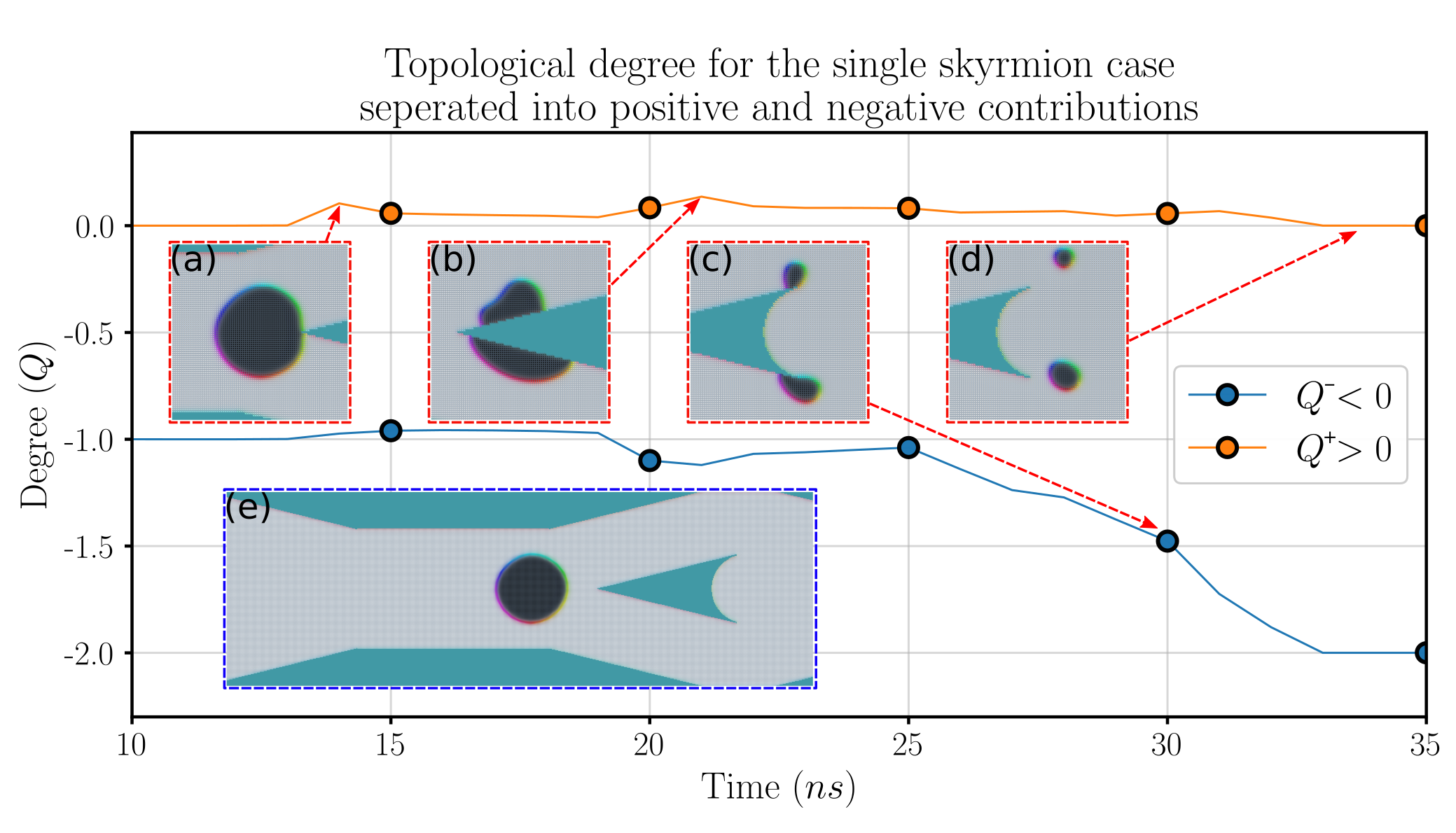}
\caption{\textbf{Cutting a skyrmion into two skyrmions.} \textit{ The graph shows the topological degree, Eq. (\ref{TopDeg}), separated into positive and negative contributions over time. The insets mark the points in time at which some critical changes occur. $Q^{+} = \frac{1}{2} \int (|q(x)|+q(x)) d^2x$ and $Q^{-} = \frac{1}{2} \int (|q(x)|-q(x)) d^2x$.
   \textbf{(a)} A simulated $S(0)$ skyrmion encountering the cutting surface. 
   \textbf{(b)} The skyrmion is completely cut in two.
   \textbf{(c)} The two N\'eel domain walls on the boundary reach the end of the cutting surface.
   \textbf{(d)} Two skyrmions.
   \textbf{(e)} The entire simulated cutting space hosting a single skyrmion prior to cutting.
  }}
\label{fig3}
\end{figure}

The form of the skyrmion slicing domain, Fig. \ref{fig3}(e), is the first one we considered. Successful cutting of bags and individual skyrmions was achieved with minimal adjustments to the timing and magnitude of the applied current. Included are two examples, additional examples can be found in the SI.

To divide a single skyrmion we apply a external magnetic field of $130$ mT in the $-e_z$ direction in order to increase the size of the skyrmion. We then apply a $40$ nA$\:$nm$^{-2}$ current and push the skyrmion onto the slicing surface which can be considered a void or non-magnetic area of the sample. When the split skyrmion reaches the end of the slicing surface a jolt of $80$ nA$\:$nm$^{-2}$, for $2$ ns, is applied in order to separate the domain wall from the surface and reform the skyrmions. The external field is switched off and the skyrmions are then pushed by current for a short time and then allowed to rest. See Fig. \ref{fig3}.


To divide a skyrmion bag we follow a similar process. The bag to be sliced contains textures of opposite degree. By applying a current sufficient to move the inner textures of a skyrmion bag apart but low enough to minimise stripe distortion, we push the bag onto the cutting surface. Here no external magnetic field is required. We apply a $20$ nA$\:$nm$^{-2}$ current and push the skyrmion bag onto the slicing surface. A larger current of $40$ nA$\:$nm$^{-2}$, for $5$ ns, is required to pierce the back wall of the skyrmion bag. When the split bag reaches the end of the slicing surface a jolt of $80$ nA$\:$nm$^{-2}$, for $1$ ns, is applied in order to separate the lower domain wall from the surface. The upper bag separated cleanly. The bags are then pushed by current for a short time and then allowed to rest.

\begin{figure}[h!]
\centering
\includegraphics[width = 0.95\columnwidth]{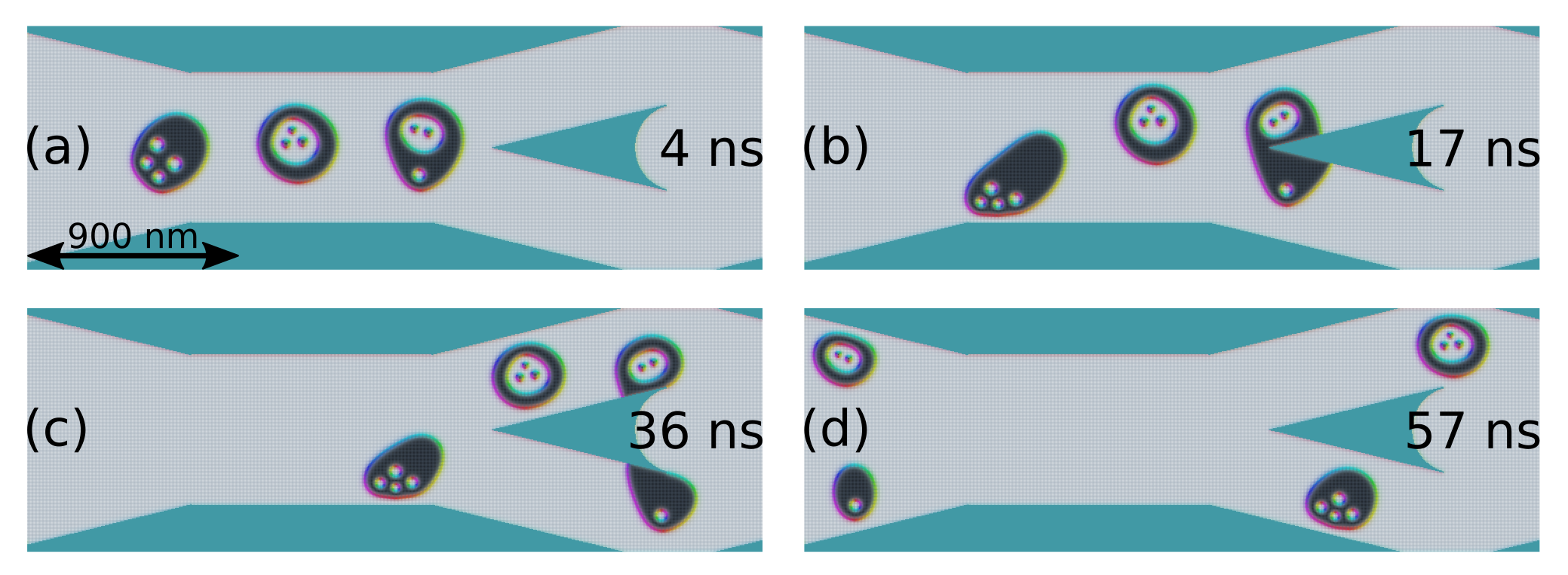}
\caption{\textbf{Cutting a skyrmion bag into two skyrmion bags.} \textit{ An $S(4)$, $S(\tilde{S}(2))$ and $S(\tilde{S}(2),\tilde{S}(0))$ in the cutting space. Their respective topological degrees are $Q=3$, $Q=-3$ and $Q=-1$. Applied current will cause the first two bags paths to diverge and they will travel below and above the cutting surface respectively.
   \textbf{(a)} $4$ ns after current is first applied to the system.  
   \textbf{(b)} The $S(\tilde{S}(2),\tilde{S}(0))$ front boundary is penetrated.
   \textbf{(c)} The N\'eel outer boundaries of the separated bags approach the end of the cutting surface.
   \textbf{(d)} The complete $S(\tilde{S}(2))$ and $S(\tilde{S}(0))$ are separated whilst the $S(4)$, and $S(\tilde{S}(2))$ pass either side of the cutting surface.
  }}
\label{fig4}
\end{figure}

The skyrmion bag slice simulation included herein contains additional skyrmion bags with opposing topological degrees which each move down a different path around the cutting surface illustrating binning \cite{kindMagneticSkyrmionBinning2021} as shown in Fig. \ref{fig4}.


\section{Discussion}

Skyrmionic division provides an additional degree of freedom to skyrmionic systems utilising both individual and composite skyrmions. Having successfully divided skyrmions and skyrmion bags in simulation we have no reason to doubt that this will be possible physically. The timings of the required current applications are tight in this initial model but further work will improve the system geometry and parameter configuration and lessen the constraints.

In the cases presented a current pulse was required to cleanly separate some of the domain walls from the slicing surface. This may be an artefact of the chosen geometry of the slicer or a feature of this system, further research will be required to analyse optimal cutting geometries and system regimes and extend skyrmionic theory.


This result adds a powerful new tool to the spintronics arsenal and emphasizes the fundamental differences between CMOS technology and skyrmionics. Skyrmionic systems can be constructed as binary storage and basic logic gates or treated as fundamentally novel technology with attributes akin to those seen in particle interactions and replicating bio-like systems. This higher dimensionality of skyrmionic systems allows for a greater range of possible devices as this result exemplifies. Alongside applications in data storage and logic devices there are therefore potential uses in the cutting edge fields of unconventional computing \cite{zieglerNovelHardwareConcepts2020}, neural modelling \cite{songSkyrmionbasedArtificialSynapses2020} and artificial life \cite{aguilarPresentFutureArtificial2014}.

\section{Appendix}

We consider skyrmionic structures in thin film multilayers and the interfacial DMI results in N\'eel type skyrmions, see Fig. \ref{fig1}. We draw our parameter regime from the range where skyrmion bags are expected to be stable \cite{kindExistenceStabilitySkyrmion2020}. 

The simulations where performed using the GPU-accelerated micromagnetic simulation program MuMax3 \cite{vansteenkistedesignverificationMuMax32014} with Landau-Lifshitz dynamics in the form
\begin{align}\label{llg}
\frac{\partial \bn}{\partial t}=\gamma\frac{1}{1+\alpha^2}\left(\bn \times \bB_{\mathrm{eff}}+\alpha\bn
\times\left( \bn \times \bB_{\mathrm{eff}} \right) \right),
\end{align}
where $\gamma \approx 176$ rad(ns$\:$T)$^{-1}$ is the electron gyromagnetic ratio, $\alpha=0.3$ the dimensionless damping parameter, $\bB_{\mathrm{eff}} = \delta E / \delta \bn$ the effective field and $\bn(\bx)=\bN (\bx)/M_{s}$ the magnetisation vector field normalised by the saturation magnetisation. The initial bag configurations were built from template functions of individual skyrmions and UD-skyrmions. The simulations showed a significant drop in energy from the initial condition which indicates the system achieving stability.

The simulation geometry is typically $(3\times1024)\times1024$ nm$^2$ rectangle of $1$ nm thickness, in order to represent a typical wire that could be fabricated using lithographic processing. Cell size of $2 \times 2 \times 1$ nm$^3$ has been used. Material parameters are: saturation magnetisation $M_{s} = 580$ kAm$^{-1}$, exchange $20$ pJm$^{-1}$, interfacial DMI $3$ mJm$^{-2}$ and uniaxial anisotropy along the $+z$ direction $0.53$ MJm$^{-3}$. 

The applied current is of Slonczewski (CPP) type,

\begin{align}\label{Slonc}
\overrightarrow{\tau}_{SL} &= \beta \frac{\epsilon - \alpha \epsilon '}{1 + \alpha^2} \left( \bn \times \left( \bn_p \times \bn \right) \right) \nonumber \\
&- \beta \frac{\epsilon ' - \alpha \epsilon}{1 + \alpha^2} \bn \times \bn_p \\
\beta &= \frac{j_z \hbar}{M_s e d} \\
\epsilon &= \frac{P \left( \overrightarrow{\br} , t \right) \Lambda^2}{\left( \Lambda^2 + 1 \right) + \left( \Lambda^2 - 1 \right)  \left( \bn \cdot \bn_p \right)} 
\end{align}

where $j_z$ is the current density along the $z$-axis, $d$ the free layer thickness, $\bn_p = -\be_y$ the fixed layer magnetisation, $P=0.4$ the spin polarisation with $\Lambda = 1$ the Slonczewski parameter and $\epsilon ' = 0$ the secondary spin torque parameter.

\section*{Acknowledgements}
This work was supported by the Engineering and Physical Sciences Research Council (EPSRC) grant EP/M506473/1. 
The Titan V GPU used for parts of this research was donated by the NVIDIA Corporation.

\bibliography{skyrmslice}

\section{Supplemental Information}
\subsection{Hedgehog ansatz, derived energy and Derrick scaling}
The configuration of a single axially symmetric skyrmion may be understood using the so-called \emph{hedgehog ansatz} \cite{nagaosaTopologicalPropertiesDynamics2013} which separates radial $r$ and angular $\theta$ coordinates with respect to the skyrmion's centre, 
\begin{align} \label{anz}
\bn(\bx)=\left(\sin f(r)\cos(m\theta+\gamma),\sin f(r)\sin(m\theta+\gamma),\cos f(r) \right),
\end{align}
where $r^2=x^2+y^2$, $f(r)$ is some monotonic function satisfying $f(0)=\pi$, $f(\infty)=0$, $\theta$ is the polar angle of $\bx$, $m\in \mathbb{Z}$ is the topological degree, and $\gamma$ is the internal orientation of the skyrmion (sometimes called the chirality).

It has been shown that the minimum energy configuration for a single skyrmion or a skyrmion bag is for the outer edge to be axially symmetric \cite{melcherChiralSkyrmionsPlane2014}. The axially symmetric skyrmion minimises over all homotopy classes with energy \cite{fosterTwodimensionalSkyrmionBags2019},

\begin{multline}\label{engrA}
E=2\pi \int \left( \frac{1}{2}f'^2(r)+\frac{m^2}{2r^2}\sin^2 f(r) + \right. \\  \left. D\frac{\sin(m\pi)\sin(\gamma+m\pi)}{m-1}\left(f'+\frac{m}{2r}\sin \left(2 f(r)\right)\right) -\mu \cos f(r)\right) r dr.
\end{multline}

Under the Derrick scaling \cite{derrickCommentsNonlinearWave1964} argument, $r \to \lambda r$, it can be shown that $m=1$ is the only stable solution. $m=-1$ would correspond to the reverse order of colour as depicted in this paper using the Runge colour sphere.

\begin{figure}[h!]
\centering
\includegraphics[width = 0.95\columnwidth]{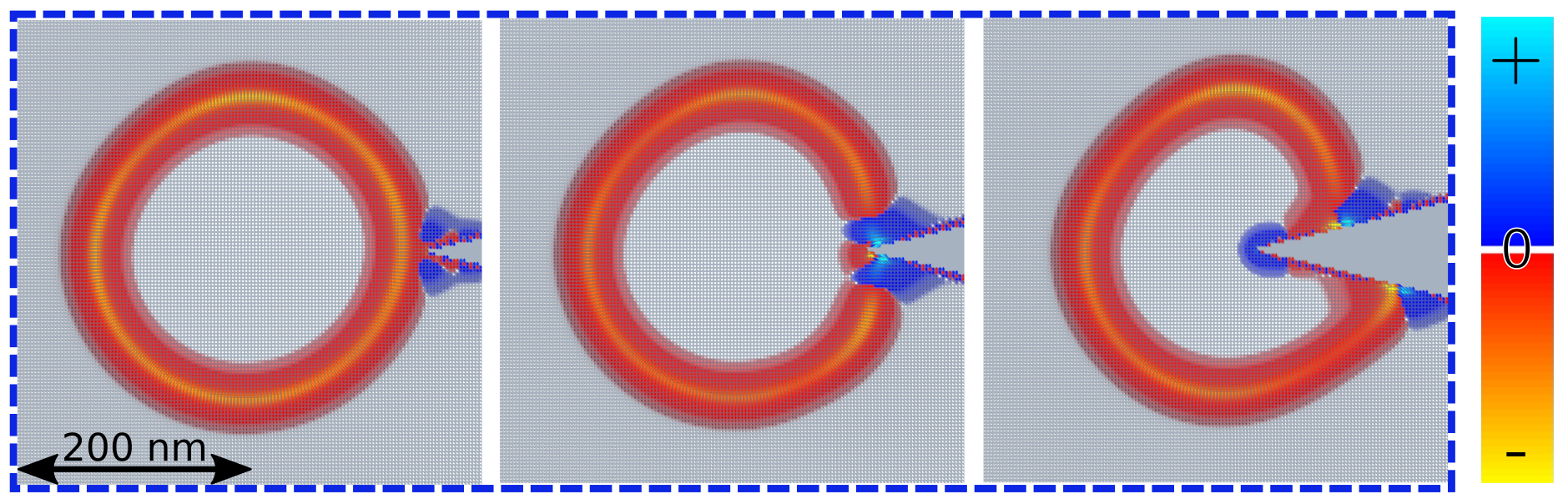}
\caption{\textbf{Negative density at the cutting edge.}
   \textit{Here a single, $S(0)$, skyrmion is pushed against the cutting edge. Note the formation of the areas of negative $q(x)$.
  }}
\label{fig5}
\end{figure}

As the skyrmion bag, or skyrmion, is deflected by the cutting defect a region of space is created with opposite topological degree. This can be seen as a region effectively encapsulated by colours of the opposite order to the globally stable skyrmion. When the skyrmionic structure is further pushed onto the defect this region of negative topological degree grows, see Fig. \ref{fig5}. Hence as the skyrmionic structure deforms it creates a region of negative topological degree until it becomes energetically favourable for the skyrmionic structure to separate into two skyrmionic structures and a third region of negative topological degree which, as shown by equation (\ref{engrA}), shrinks to a point.

\begin{align} \label{Einequ}
E_{S(n)} + E_{S(m)} + E_{negative skyrmion} < E_{deformed S(n+m)},
\end{align}

where $E_{S(n)}$ and $E_{S(m)}$ are the energies of the successfully cut skyrmionic structure, $E_{negative skyrmion}$ is the energy of the opposing degree skyrmion and $E_{deformed S(n+m)}$ is the total energy of the deformed skyrmionic structure.

\subsection{Two more cutting examples}

\begin{figure}[h!]
\centering
\includegraphics[width = 0.95\columnwidth]{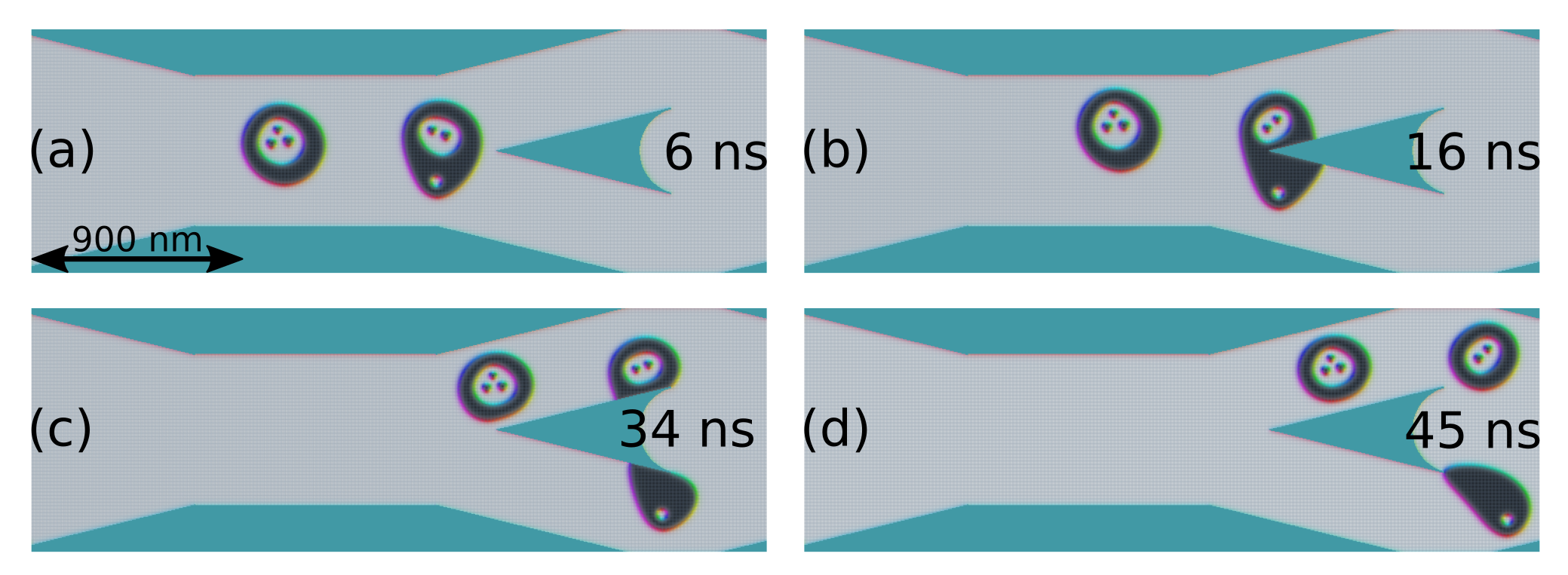}
\caption{\textbf{Cutting an $S(\tilde{S}(2),\tilde{S}(0))$.}
   \textit{
   \textbf{(a)} $6$ ns after current is first applied to the system.  
   \textbf{(b)} The $S(\tilde{S}(2),\tilde{S}(0))$ front boundary is penetrated.
   \textbf{(c)} The N\'eel outer boundaries of the separated bags approach the end of the cutting surface.
   \textbf{(d)} The complete $S(\tilde{S}(2))$ and $S(\tilde{S}(0))$ separate from the cutting surface whilst the $S(\tilde{S}(3))$ passes above the cutting surface.
  }}
\label{fig6}
\end{figure}

\begin{figure}[h!]
\centering
\includegraphics[width = 0.95\columnwidth]{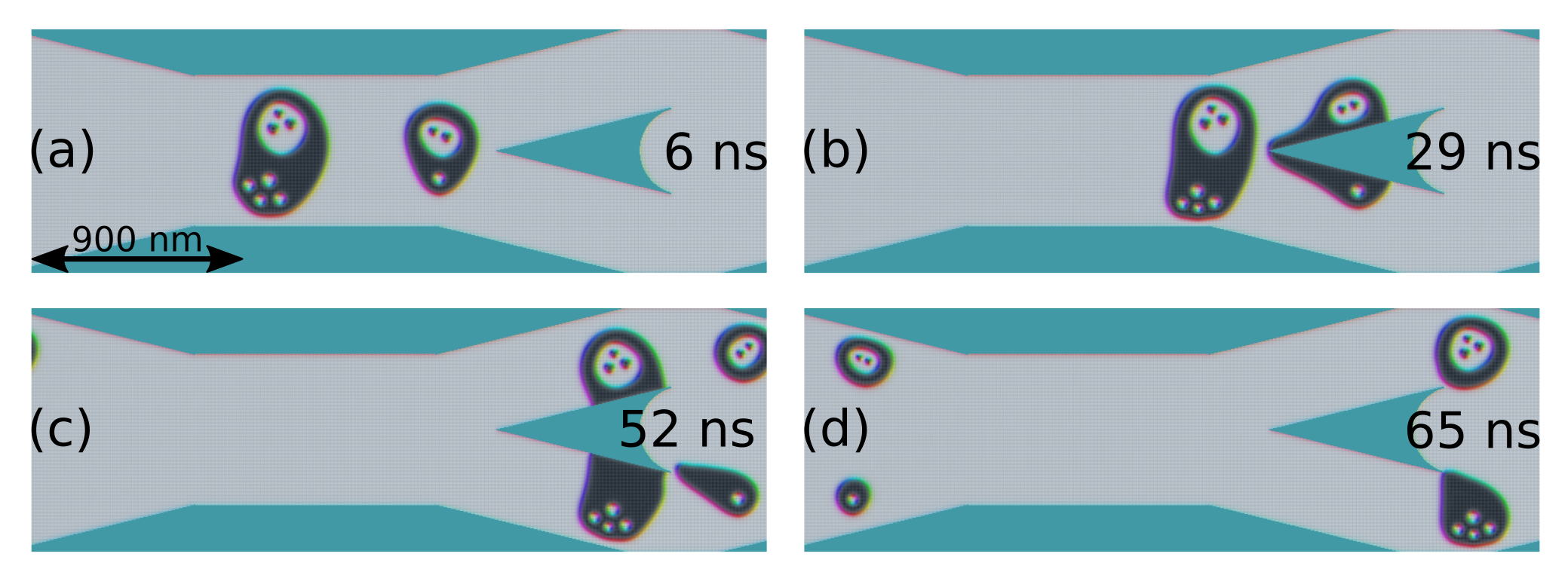}
\caption{\textbf{Cutting an $S(\tilde{S}(2),\tilde{S}(0))$ and an $S(\tilde{S}(3),4 \tilde{S}(0))$.}
   \textit{
   \textbf{(a)} $6$ ns after current is first applied to the system.  
   \textbf{(b)} The $S(\tilde{S}(2),\tilde{S}(0))$ is half way along the cutting surface and the $S(\tilde{S}(3),4 \tilde{S}(0))$ approaches the cutting edge.
   \textbf{(c)} The $S(\tilde{S}(2),\tilde{S}(0))$ has separated and the $S(\tilde{S}(3),4 \tilde{S}(0))$ is nearing the end of the cutting surface.
   \textbf{(d)} The complete $S(\tilde{S}(2))$ and $S(\tilde{S}(0))$ are separated whilst the $S(\tilde{S}(3))$ and $S(4)$ are leaving the cutting surface.
  }}
\label{fig7}
\end{figure}

\end{document}